\documentclass[a4paper,twocolumn,prb]{revtex4-2}
\usepackage[english]{babel}
\usepackage[utf8]{inputenc}
\usepackage{amsmath}
\usepackage{amsfonts}
\usepackage{mathtools}

\usepackage{bm}
\usepackage{siunitx}
\usepackage{xspace}
\usepackage{microtype}
\usepackage{tabularx}
\usepackage{makecell}

\newcommand{\norm}[1]{\lvert #1 \rvert}
\newcommand{\braket}[1]{\langle #1 \rangle}
\newcommand{\ket}[1]{\vert #1 \rangle}
\newcommand{\bra}[1]{\langle #1 \vert}
\newcommand{\mel}[3]{\langle #1 \vert #2 \vert #3 \rangle}

\renewcommand{\dag}{^\dagger}
\DeclareMathOperator{\Tr}{Tr}
\DeclareMathOperator{\hc}{H.c.}
\renewcommand{\phi}{\varphi}

\usepackage{bbold}
\newcommand{\id}{\mathbb{1}}

\usepackage[colorlinks=true,linkcolor=black,citecolor=black]{hyperref}
\usepackage[capitalise]{cleveref}

\usepackage{graphicx}
\graphicspath{ {figures/} }

\usepackage{natbib}

\newcommand{\affA}{Department of Physics and Astronomy, Aarhus University, DK-8000 Aarhus C, Denmark}
\newcommand{\affB}{Department of Chemistry, Aarhus University, DK-8000 Aarhus C, Denmark}

\date{\today}
\begin{document}
	\title{Dynamical quantum phase transitions in a noisy lattice gauge theory}
	
	\author{Rasmus Berg Jensen}
	\email{rbj@chem.au.dk}
	\affiliation{\affA}
	\affiliation{\affB}
	\author{Simon Panyella Pedersen}
	\email{spp@phys.au.dk}
	\affiliation{\affA}
	\author{Nikolaj Thomas Zinner}
	\email{zinner@phys.au.dk}
	\affiliation{\affA}

\begin{abstract}
	Lattice gauge theories (LGTs) form an intriguing class of theories highly relevant to both high-energy particle physics and low-energy condensed matter physics with the rapid development of engineered quantum devices providing new tools to study e.g. dynamics of such theories. The massive Schwinger model is known to exhibit intricate properties of more complicated theories and has recently been shown to undergo dynamical quantum phase transitions out of equilibrium. With current technology, noise is inevitable and potentially fatal for a successful quantum simulation. This paper studies the dynamics subject to noise of a $(1+1)$D U$(1)$ quantum link model following a quench of the sign of the mass term. We find that not only is the system capable of handling noise at rates realistic in NISQ-era devices, promising the possiblity to study the target dynamics with current technology, but the effect of noise can be understood in terms of simple models. Specifically the gauge-breaking nature of bit-flip channels results in exponential dampening of state amplitudes, and thus observables, which does not affect the structures of interest. This is especially important as it demonstrates that the gauge theory can be successfully studied with devices that only exhibit approximate gauge invariance.
\end{abstract}

\maketitle

\section{Introduction}\label{sec:intro}
In recent years the possibilities for experimentally studying challenging avenues of physics through engineered quantum simulators, i.e. experiments designed to emulate otherwise inaccessible physics, have increased rapidly. Such studies, called analog quantum simulations, is currently recieving great experimental and theoretical attention as many challenges remain from both sides \cite{georgescuQuantumSimulation2014,bondEffectMicromotionLocal2022,yingFloquetPrethermalPhase2022,huergaVariationalQuantumSimulation2022,viermannQuantumFieldSimulator2022}. Specifically gauge theories have been the target of much attention \cite{banulsSimulatingLatticeGauge2020} since these are notoriously challenging --- both analytically and numerically. Due to their importance for our understanding of Nature, gauge theories are thus very interesting targets for analog quantum simulations.

Originally invented to study quark confinement in quantum chromodynamics \cite{wilsonConfinementQuarks1974} lattice gauge theories (LGTs) are currently experiencing renewed attention with their potential for numerical and quantum simulation of systems relevant to both high-energy particle physics \cite{katoDoublewindingWilsonLoops2020,shibataLatticeStudyArea2018,zoharQuantumSimulationLattice2021,halimehEnhancingDisorderfreeLocalization2021,bernienProbingManybodyDynamics2017} and low-energy condensed matter physics \cite{banerjeeInterfacesStringsSoft2014,islamMeasuringEntanglementEntropy2015, greifShortRangeQuantumMagnetism2013}. Such simulations have the potential to enlighten the otherwise scarcely known non-equilibrium structure of LGTs. This can be studied through dynamical quantum phase transitions (DQPTs) that occur at zeros or other critical behaviour of the Loschmidt amplitude,
\begin{align}
    \mathcal{G}(t) = \braket{\psi(0) \vert \psi(t)},
\end{align}
which is the overlap between a state at time $t$ and the initial state it evolved from. Analogous to equilibrium phase transitions, signified by critical behaviour in the partition function, DQPTs reveal changes in the physical properties of a system. A class of LGTs especially suited for the study of DQPTs through analog quantum simulation is quantum link models (QLMs) mapping highly interesting LGTs to spin models. It has recently been shown that even the non-equilibrium dynamics of QLMs exhibit good convergence towards the  Wilson–Kogut–Susskind limit \cite{halimehAchievingQuantumField2021}.

One of the primary issues with quantum simulations in the era of noisy intermediate-scale quantum (NISQ) devices is handling the noise inevitably present in the simulation. Specific to LGTs is the issue of retaining the gauge symmetry which defines the target model --- currently a topic of active study \cite{halimehEnhancingDisorderfreeLocalization2021,halimehGaugeProtectionNonAbelian2021,halimehStabilizingDisorderFreeLocalization2021,halimehStabilizingLatticeGauge2021,mazzolaGaugeinvariantQuantumCircuits2021,halimehFateLatticeGauge2020,vandammeGaugeSymmetryViolationQuantum2020}. There are studies focusing on reducing the impact of noise in quantum simulations of LGTs \cite{gustafsonNoiseImprovementsQuantum2022,funckeQuantumSimulationsParticle2022,yamamotoRealtimeSimulationDimensional2021}, but to the best of our knowledge it has not been studied how noise affects DQPTs in realistic quantum simulations. Recently DQPTs in a QLM version of the massive Schwinger model \cite{schwingerGaugeInvarianceMass1962,colemanChargeShieldingQuark1975,colemanMoreMassiveSchwinger1976} were studied along with a novel order parameter whose zeros correlate with minima in the Loschmidt echo \cite{pedersenLatticeGaugeTheory2021}, $\mathcal{L}(t) = |\mathcal{G}(t)|^2$. However, the impact of noise was not studied there and this is what we address here.

\Cref{sec:system} introduces the system that forms the topic of the study and defines the necessary observables in the open quantum systems formalism. The noise models used are introduced in \cref{sec:sim} which also contains the results of numerical simulations. These simulations show the effect of the two types of noise channels on the target structures along with a simpler models aiding the understanding of the noise affected dynamics. The results of similar investigations of the order parameter is shown before the target correlation between order parameter zeros and Loschmidt echo minima, and thereby also DQPTs, is studied. It is shown that the system is perfectly capable of handling noise at relaxation rates corresponding to experimentally oberserved lifetimes of NISQ-era devices, thereby showing that the intricate non-equilibrium dynamics of the model can realistically be studied with existing quantum technology. In \cref{sec:scaling} longer spin chains are considered showing that scaling the system to larger sizes does not seem to introduce any unexpected noise effects. The previous conclusion should thus be valid for systems of increasing size. Lastly the paper is summarized and conclusions are drawn in \cref{sec:conclusion}.

\section{System and procedure}\label{sec:system}
The system of interest is an interacting LGT, specifically the $(1+1)$D U$(1)$ gauge theory, on a periodic lattice, which is interesting as the dynamics probe the scarely known non-equilibrium structure of the model with DQPTs providing a lense through which to conduct the study --- a topic that has therefore attracted recent attention in the litterature \cite{magnificoRealTimeDynamics2020,nguyenDigitalQuantumSimulation2021}.

Let $\psi_n$ be the staggered mass ($m$) fermion operator at lattice site $n$, $U_{n,n+1}$ and $E_{n,n+1}$ the link and electric field operators respectively between lattice sites $n$ and $n+1$ and $a$ the lattice spacing. The Hamiltonian of the model can then be expressed as \cite{zacheDynamicalTopologicalTransitions2019}
\begin{align} \label{eq:lgt_hamiltonian}
\begin{aligned}
    H_\textsc{lgt} = a \sum_n \bigg[ &(-1)^{n}m\psi\dag_n\psi_n + \frac{1}{2} E_{n,n+1}^2 \\
    &- \frac{i}{2a} \left( \psi\dag_n U_{n,n+1} \psi_{n+1} + \hc \right) 
    \bigg].
\end{aligned}
\end{align}
The LGT Hamiltonian is mapped to a quantum link model (QLM) of coupled spin-1/2 systems by representing the fermionic field as Jordan-Wigner transformed staggered mass fermions and the gauge fields by spin-1/2 degrees of freedom.
Let $N$ denote the number of matter sites and $J = 1/a$ the matter-gauge coupling constant. The system is then described by the Hamiltonian \cite{pedersenLatticeGaugeTheory2021}
\begin{align}
	H = \sum_{n=0}^{N-1} \left[ -(-1)^{n}\frac{m}{2}\sigma_{n}^{z} + \frac{J}{2} \Big( \sigma_{n}^{+}S_{n,n+1}^{+}\sigma_{n+1}^{-} + \hc \Big) \right]. \label{eq:H}
\end{align}
Here $\sigma^\alpha_n$, with $\alpha = z, \pm$, denotes the usual spin-1/2 operators pertaining to matter site $n$ represented in this case by the Pauli-Z and step-up/down matrices. Similarly $S_{n,n+1}^\alpha$ denotes the spin operator on the link between lattice sites $n$ and $n+1$. Due to the spin-1/2 representation the electric field kinetic energy term is constant and therefore neglected. The Hamiltonian is expressed in terms of the coupling constant $J$ rather than the lattice spacing $a$, since the continuum limit is not of concern for this study. Representing the gauge link degrees of freedom makes the model ideal for quantum simulator even though the model differs somewhat from the Wilon-Kogut-Susskind limit \cite{vandammeDynamicalQuantumPhase2022}.

The target dynamics is achieved by quenching the sign of the mass, $m \rightarrow -m$ at time $t = 0$. That is the system is initialized in the ground state of the Hamiltonian in \cref{eq:H}, $H = H(m,J)$, and time-evolution is then performed according to the Hamiltonian with the mass of the opposite sign, $H(-m,J)$. The unitary time-evolution of the system following this procedure and its implementation using NISQ-era devices was the topic of Ref. \cite{pedersenLatticeGaugeTheory2021}, but that paper left the impact of noise on the simulation unexplored which this paper addresses.

By construction, \cref{eq:H} obeys a local U$(1)$ gauge symmetry generated by
\begin{align}
	G_{n} = S_{n-1,n}^{z} - S_{n,n+1}^{z} + \sigma_{n}^{z} - (-1)^{n},
\end{align}
which partitions the Hilbert space, $\mathcal{H}$, in gauge sectors identified by the eigenvalues of $G_n$ denoted $\{g_n\}$. Specifically the system is initialized in the gauge sector where $g_n = 0 \; \forall n$. As the gauge symmetry is conserved across the quench, the target dynamics is confined to the $g_n = 0$ sector which substantially reduces the dimension of the effective Hilbert space. Since $\dim\mathcal{H} = 2^{2N}$ for the full Hilbert space of \cref{eq:H} it is vital to the scalability of any numerical simulation to confine the simulation to the effective Hilbert space. Noise processes can, however, not be expected to conserve gauge symmetry. Hence, simulations of the non-unitary dynamics cannot generally be confined to the initial gauge sector as in the unitary case, severely hampering the scalability of the simulation with respect to the system size. For this reason unless otherwise stated this study is concerned with the $N = 4$ case, which is large enough to exhibit the structures of interest \cite{pedersenLatticeGaugeTheory2021}, but small enough for detailed simulations.

\subsection{Observables}
In the case of unitary time-evolution the observables of interest for this study are the Loschmidt amplitude, $\mathcal{G}(t) = \braket{\psi(0) \vert \psi(t)}$, the Loschmidt echo, $\mathcal{L} = \norm{\mathcal{G}}^2$, and a specific order parameter introduced by Ref. \cite{pedersenLatticeGaugeTheory2021}. The order parameter $g(k,t)$ is the discrete Fourier transform of the sum of amplitudes for processes where a matter site excitation moves from either site $0$ or $1$ to site $n$ between initialization and time $t$. Both sites $0$ and $1$ are used such that it is symmetric with respect to particles and antiparticles. Defining the operator
\begin{align} \label{eq:order_param_op}
	\mathfrak{g}(k) = \sum_{m=0,1} \sum_{\mathclap{n=0}}^{\mathclap{N-1}} e^{-ikd_{m}(n)} \sigma_{m}^{-} \prod_{i=m}^{n-1} S_{i,i+1}^{\alpha_{m}(n)}\sigma_{n}^{+},
\end{align}
the order parameter is defined as the matrix element
\begin{align} \label{eq:order_param}
	g(k,t) = \mel{\psi(0)}{\mathfrak{g}(k)}{\psi(t)}.
\end{align}
Here $d_m(n)$ is the shortest distance between lattices sites $n$ and $m$, where clockwise paths are measured as positive and counter-clockwise paths negative (recall that periodic boundary conditions are assumed). To ensure gauge invariance, the gauge link operator is therefore path dependent with $\alpha_{m}(n) = -$ for clockwise paths and for counter-clockwise paths $\alpha_{m}(n) = +$. For site pairs on opposite sites of the periodic chain where both distances are equal in length, i.e. $m-n = N/2$, both paths are included in the sum. As $g(k,t)$ is symmetric in $k$, due to the periodic boundary conditions, it suffices to consider half of the Brillouin zone, i.e. $k \in [0,\pi/a]$ where $a = 1/J$ is the lattice spacing. For a more detailed discussion of the order parameter see \cite{pedersenLatticeGaugeTheory2021}.

Ref. \cite{pedersenLatticeGaugeTheory2021} showed that the zeros of the order parameter correlate with troughs of the Loschmidt echo and that the phase of the order parameter, defined as $\phi_g$ such that $g = |g|\exp(i\phi_g)$, exhibit non-trivial vortex dynamics in the $Jt$-plane. It is therefore of interest to study how this order parameter responds to noise.

\subsubsection{Non-unitary case} \label{sec:obs_noise}
To investigate the response of the system to noise, the observables of interest have to be generalized to the open quantum systems formalism used to simulate the noise-affected dynamics. Let $\rho(t)$ denote the density matrix of the system at time $t$, which evolves in time according to the Lindblad master equation. The Loschmidt echo becomes the Frobenius inner product of the two density matrices
\begin{align} \label{eq:loschmidt_echo_noise}
	\mathcal{L}(t) = \Tr \rho\dag(0) \rho(t).
\end{align}
The Loschmidt amplitude is non-trivial to generalize and beyond the scope of this paper. It has been shown to be theoretically possible \cite{uhlmannBerryPhasesMixtures1989,uhlmannParallelTransportQuantum1986},  and an approach for experimental observation based on interferometry has been proposed \cite{sjoqvistGeometricPhasesMixed2000} and used \cite{ghoshExperimentalMeasurementMixed2006}. These works were however focussed on unitary time-evolution of a mixed state --- not the non-unitary dynamics of an initially pure state and the Loschmidt phase can therefore not be calculated directly. The generalization of the order parameter is also challenging since in the pure state limit
\begin{align}
	\Tr \rho\dag(0) \mathfrak{g}(k) \rho(t) \xrightarrow{\rho \rightarrow \ket{\psi}\bra{\psi}} g(k,t) \mathcal{G}^*(t).
\end{align}
Hence, in the open quantum systems formalism the order parameter phase includes the unknown Loschmidt phase. To account for the norm of the Loschmidt amplitude the order parameter studied here is
\begin{align} \label{eq:order_param_noise}
	g(k,t) = \frac{\Tr \rho\dag(0) \mathfrak{g}(k) \rho(t)}{\sqrt{\mathcal{L}(t)}}.
\end{align}
The phase of this order parameter contains contributions from both \cref{eq:order_param} and the Loschmidt phase which is sufficient for this study.

\subsubsection{Locating zeros} \label{sec:locating_zeros}
Of critical importance to the study of DQPTs and the order parameter is the ability to accurately locate zeros of a complex function of two real variables $f(x,y)$, written in the polar form as $f = |f|\exp(i\phi)$. One can construct an algorithm relying on the phase $\phi$ to locate zeros, which can reliably destinguish true zeros from points of small, but non-zero modulus --- even on coarse grids of data \cite{pedersenLatticeGaugeTheory2021,fukuiChernNumbersDiscretized2005}. This method will be elaborated and used in the study of the order parameter in \cref{sec:order_param}.

Since the Loschmidt phase becomes unknown in the noisy case (\cref{sec:obs_noise}) DQPTs are found as minima of the Loschmidt echo using a custom minimization algorithm for a discrete set of points based on the usual steepest decent method, where the zeros from the noiseless case serve as initial points.

\section{Simulation of dynamics} \label{sec:sim}
To simulate the noise-affected dynamics of the system the time-evolution is performed by numerically solving the Lindblad master equation with a noise model specified through the choice of Lindblad operators. Assuming that noise processes can be described as Markovian, the Lindblad operators can be expressed in the usual basis of spin operators as
\begin{align}
	L^\alpha_i = \sqrt{\gamma} \hat{L}^\alpha_i,
\end{align}
where $\gamma$ is the relaxation or noise rate, $i$ is an index that includes both matter sites and gauge links and
\begin{align} \label{eq:lindblad_op}
	\hat{L}^\alpha_i = \id_1,\dots{},\sigma^\alpha_i,\dots{},\id_{2N}.
\end{align}
The set of Lindblad operators is denoted $\mathfrak{L} = \{\hat{L}^\alpha_i\}$. Lindblad operators on the form of \cref{eq:lindblad_op} can be interpreted as describing two types of processes: bit-flips and phase decoherence, which flips the qubit or adds a phase to the affected qubit respectively. The impact of noise can thus be understood by studying the two types.

Paramount to the relevance of this type of study is the choice of relaxation rate. Employing the dimensionless time-variable $tm$ an experimental $1/e$ lifetime, $T$, can be mapped to a relaxation rate in units convenient to the simulations as
\begin{align}
	\gamma = \frac{1}{Tm},
\end{align}
which is implementation dependent through $m$. Depending on the physical system used in an actual experiment, lifetimes vary between orders of magnitude \SI{10}{\micro\second} and \SI{10}{\second} \cite{krantzQuantumEngineerGuide2019,nguyenHighCoherenceFluxoniumQubit2019,
yanFluxQubitRevisited2016,gyenisExperimentalRealizationProtected2021,wangPracticalQuantumComputers2022,
linSecondsscaleCoherenceNuclear2021,wangSingleIonQubit2021,bollinger1991,fiskVeryHighQMicroscope1995,
lawrieSpinRelaxationBenchmarks2020,onizhukProbingCoherenceSolidState2021,
watsonAtomicallyEngineeredElectron2017,herbschlebUltralongCoherenceTimes2019,
levineHighFidelityControlEntanglement2018,wilsonTrappingAlkalineEarth2022,
ya-litiantianyaliComparisonSingleneutralatomQubit2019,hermann-aviglianoLongCoherenceTimes2014} with gate times on the order of magnitude of \SI{10}{\nano\second} to \SI{1}{\micro\second} \cite{kjaergaardSuperconductingQubits2020,schaferFastQuantumLogic2018,hegdeEfficientQuantumGates2020,zhangSubmicrosecondEntanglingGate2020}. In realistic experiments one can therefore expect that $m \gtrsim \SI{1}{\mega\hertz}$. We therefore choose $T = \SI{20}{\micro\second}$ and $m = \SI{6.4}{\times 2\pi\;\mega\hertz}$ (Ref. \cite{pedersenLatticeGaugeTheory2021} found this value optimal for their proposed circuit) and thus $\gamma = \num{1.25e-3}$ in order to resemble an experimental setup with a somewhat poor lifetime to gate time ratio.

\begin{figure*}
	\includegraphics[width=\textwidth]{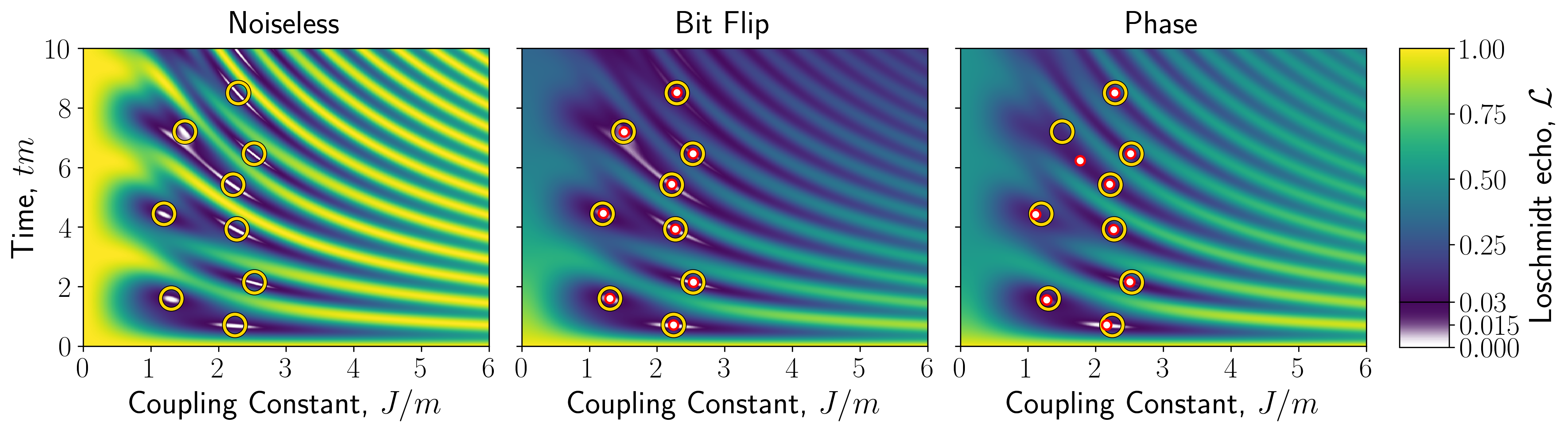}
	\caption{Loschmidt echo from simulations without noise, with bit flip noise implemented using $L^x_i = \sqrt{\gamma} \hat{L}_i^x$ and phase noise with $L^z_i = \sqrt{\gamma} \hat{L}_i^z$ in the respective panels. The relaxation rate in both panels with noise is $\gamma = \num{2.5e-2}$ and the number of matter sites is $N = 4$. Yellow circles mark the zeros of the noiseless case in all panels and the red encircled white dots indicate the minima of the displayed Loschmidt echo. To a large extend bit flip noise does not affect the structure of the Loschmidt echo --- it merely suppresses its magnitude --- making the minima remarkably robust against bit flip noise. Phase noise has a larger effect on the structure of the Loschmidt echo but its magnitude is less suppressed. The minima are thus less robust to phase flip noise than bit flip noise.}
	\label{fig:loschmidt_echo_noise}
\end{figure*}
\subsection{Relaxation Channels} \label{sec:relax}
The most basic models of the bit-flip and phase relaxation channels are
\begin{subequations} \label{eq:relax_models}
\begin{align}
	\mathfrak{L}_\textup{bit} &= \{L^x_i\}, \\
	\mathfrak{L}_\textup{phase} &= \{L^z_i\}. \label{eq:phase_model}
\end{align}
\end{subequations}
In \cref{fig:loschmidt_echo_noise}, the Loschmidt echo as a function of time and coupling constant is shown in the case of bit-flip, phase relaxation and no noise with the noise models in \cref{eq:relax_models}. Additionally, zeros from the noiseless case, identifying DQPTs, are in all panels marked with yellow circles. Red encircled white dots in the noisy cases mark Loschmidt echo minima. The relaxation rate in the noisy cases is $\gamma = \num{2.5e-2}$ chosen to illustrate how the system is affected by the noise, which is significantly larger than what is achievable in realistic experiments. Clearly the overall structure of the Loschmidt echo is preserved quite well when subject to bit-flip noise. It merely seems to suppress the magnitude of the Loschmidt echo. Phase noise on the other hand more severely tampers with the structure as seen from the displacement of the minima. The Loschmidt echo is, however, not suppressed to the same degree as for the case of bit-flip noise. As the location of a majority of the Loschmidt echo zeros can be successfully approximated by the minima even at a quite large relaxation rate, this is a promising sign that the study of DQPTs with NISQ-era devices might be possible. The study of how these structures depend on the noise rate is postponed to \cref{sec:mult_channels}.

\subsubsection{Phase decoherence} \label{sec:phase}
In more general terms across a wide range of relaxation rates one sees that the large scale structure of the function $\mathcal{L}(J/m,tm)$ is preserved quite well subject to phase relaxation. That is the wave-like pattern at $J/m \gtrsim 2$ and the broadening of troughs around $1 \lesssim J/m \lesssim 2$ can be observed to some extent at all rates explored throughout the study. However, these large scale structures become less pronounced, i.e., small values of the Loschmidt echo become larger and large values become smaller. This can be seen in \cref{fig:loschmidt_echo_noise} where all white patches apart from that at $(J/m,tm) \sim (2,1)$ have disappeared and the areas that are distinctly yellow in the noiseless case become increasingly green with time. This affects the small scale structure as seen by the fact that some minima are displaced. It is this small scale distortion that is most critical as it directly affects our ability to, e.g., locate DQPTs.

\begin{figure*}
	\includegraphics[width=\textwidth]{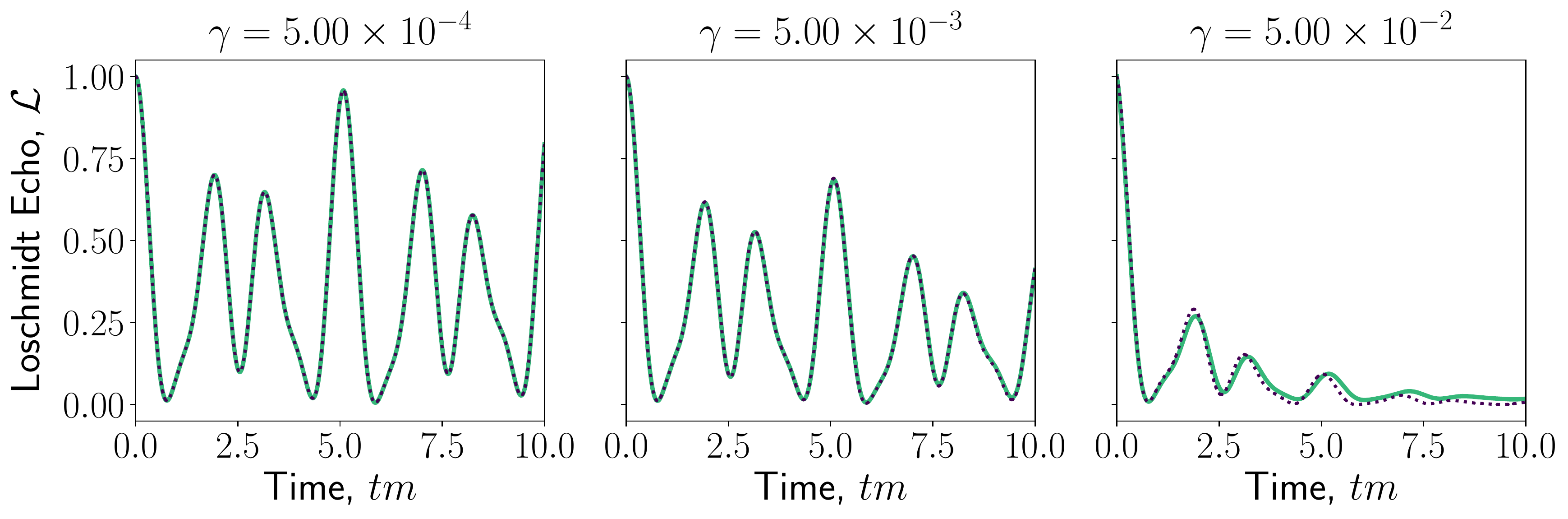}
	\caption{The Loschmidt echo, at the noise rates displayed above each panel with the coupling constant fixed at $J/m = 1.95$, is shown with a solid green line. A fit according to the conjecture of \cref{eq:bit-flip_conjecture} is displayed as a dotted purple line. In the central and left-hand panel the fit is remarkably good, whereas the right-hand panel shows some deviation from the conjecture.}
	\label{fig:loschmidt_bit}
\end{figure*}
\subsection{Bit-flip errors} \label{sec:bit_flip}
Relaxation channels that break a symmetry of the Hamiltonian, tend to induce exponential decay of amplitudes in the initial symmetry sector, e.g., spontaneous emission in atomic physics. One might therefore expect that each bit-flip channel effectively alters the Loschmidt amplitude by the factor $\exp(-\gamma t)$ which would imply that
\begin{align} \label{eq:bit-flip_conjecture}
	\mathcal{L} \xrightarrow{\textup{bit-flip}} \mathcal{L}_\textup{bit-flip} = e^{-a(\gamma) t} \mathcal{L}.
\end{align}
Here $a(\gamma) = 2n_\textup{c} \gamma$ with $n_\textup{c}$ being the number of bit-flip channels, i.e. the number of operators in $\mathfrak{L}_\textup{bit}$, since each channel contributes a factor of $\exp(\gamma t)$ to the Loschmidt amplitude and thus a factor of $\exp(2\gamma t)$ to the Loschmidt echo. As $\mathfrak{L}_\textup{bit}$ contains one operator for each spin, \cref{eq:lindblad_op}, $n_\textup{c} = 2N$. One would thus expect that $a(\gamma) = 4N\gamma$.

\cref{fig:loschmidt_bit} shows the Loschmidt echo in red and the best least-squares fit to \cref{eq:bit-flip_conjecture} with a dotted, black line. The fitted parameter is $a$ and the coupling constant is fixed at $J/m = \num{1.95}$. For the noise rates $\gamma = \num{5e-4}$ and $\num{5e-3}$ the fit is excellent and at those relaxation rates with the coupling constant fixed at $J/m = 1.95$ the conjecture certainly match the data. For the case of $\gamma = \num{5e-2}$ the fit is still reasonable, but the main difference between the simple model and the exact simulation is the location of the minima. Since this paper is concerned with the reliable identification of the Loschmidt echo zeros this deviance is troublesome. It is therefore wise to employ a quite strict limit on the range of relaxation rates where \cref{eq:bit-flip_conjecture} is valid. 

\begin{figure}
	\includegraphics[width=\columnwidth]{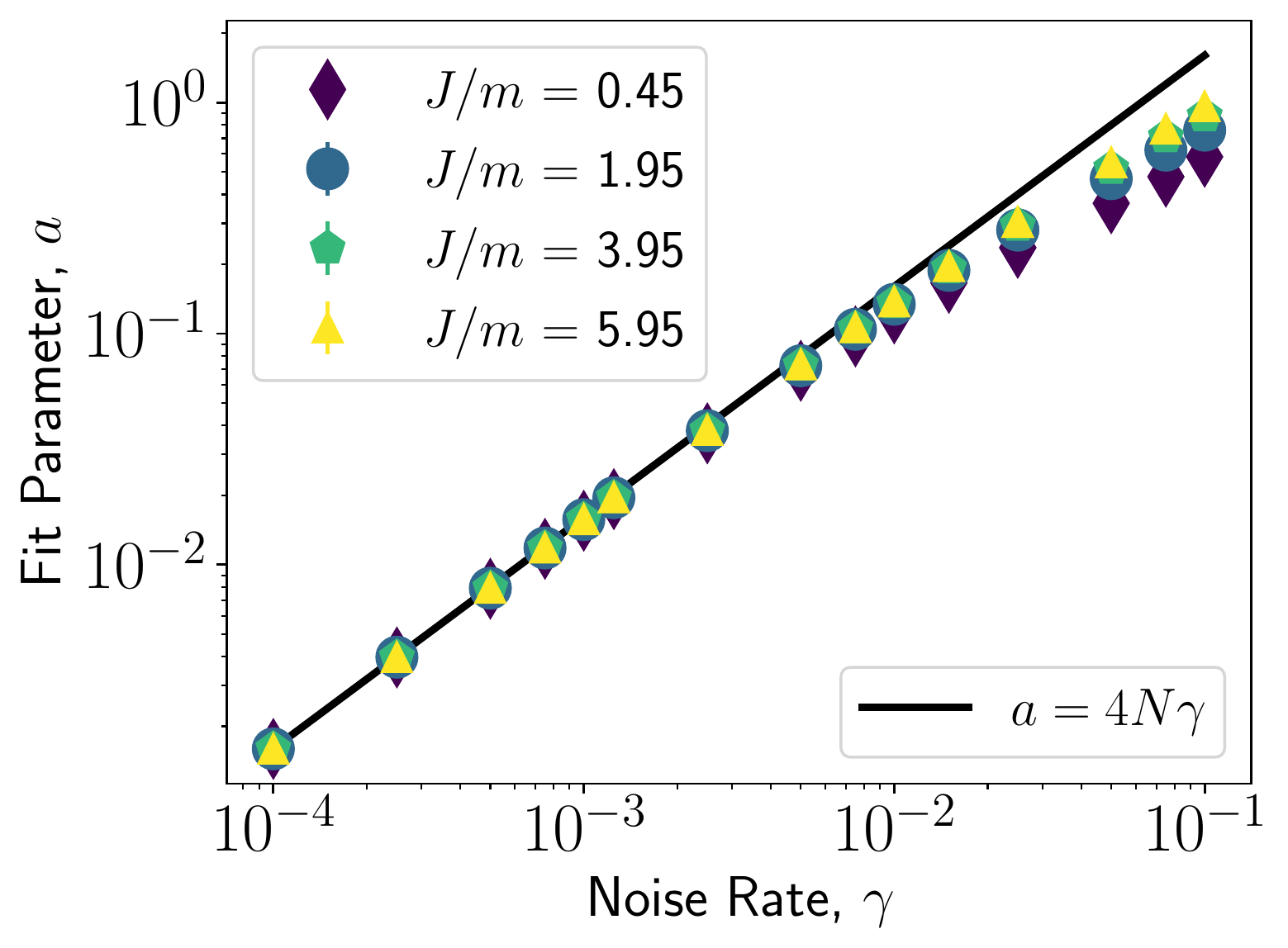}
	\caption{The model parameter of \cref{eq:bit-flip_conjecture} that yielded the best fit, like in \cref{fig:loschmidt_bit}, at a variety of noise rates and coupling constants. The black line shows the expected behaviour of the model parameter in the limit of a sufficiently small relaxation rate, $a = 4N\gamma$. The fit uncertainty is indicated by errorbars which are however too small to be visible. All data points at fixed $\gamma$ are clearly clumped together hence the model parameter is independent of the coupling constant.}
	\label{fig:bit_model_scale}
\end{figure}
The conjecture is further investigated by considering a wider range of relaxation rates and coupling constants, the result of which is seen in \cref{fig:bit_model_scale}. Along with the data a black line indicates the expected behaviour of the model parameter $a(\gamma) = 4N\gamma$.
Around $\gamma = \num{1e-2}$ the data starts to fall short of the black line. Such behaviour is indicative of a violation of the assumption on which the model is build, hence \cref{eq:bit-flip_conjecture} provides a good description of the full simulation at rates $\gamma \lesssim \num{1e-2}$.
In this range data belonging to each value of the coupling constant are indistinguishable, validating that the model is independent of $J/m$.

The excellent agreement between fit and expectation shows that \cref{eq:bit-flip_conjecture} provides not only a qualitative picture through which the bit-flip noise channels can be understood, but also an effective model that match the full simulation very well. Additionally this also implies that for $\gamma \lesssim \num{1e-2}$ bit-flip errors essentially do not impact the structure of the Loschmidt echo, hence minima do not move appreciably in the $Jt$-plane. Bit-flip error only suppress the magnitude of the Loschmidt echo and these channels are therefore not a large threat to the study of DQPTs in NISQ-era devices. 

\begin{figure*}
	\includegraphics[width=\textwidth]{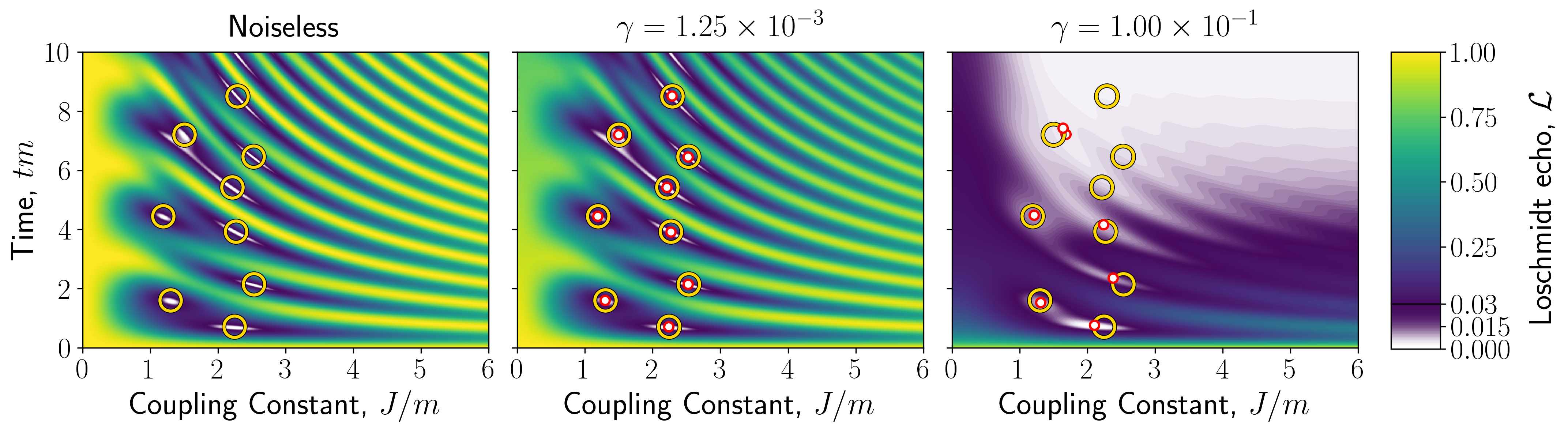}
	\caption{Loschmidt echo from simulations without noise and the noise model in \cref{eq:real_noise} at the noise rates displayed on top of the panels. In all panels the zeros from the noiseless case are marked with yellow circles and in the noisy cases minima are marked by red encircled white dots. The noise rate of the middle panel, showing a realistic case, is significantly smaller than that in \cref{fig:loschmidt_echo_noise}, hence excellent reproduction of the noiseless simulation is seen --- especially with respect to the minima. The patterns in the right-hand panel can be understood as the combined effect of the two types of noise channels (\cref{fig:loschmidt_echo_noise}) --- both suppression and distortion of the Loschmidt echo is observed.}
	\label{fig:real_noise}
\end{figure*}
\subsection{Multiple relaxation channels} \label{sec:mult_channels}
Expanding the noise model to
\begin{align} \label{eq:real_noise}
	\mathfrak{L} = \left\{\hat{L}_i^x(\gamma_\textup{phase}), \hat{L}_i^y(\gamma_\textup{phase}), \hat{L}_i^z(\gamma_\textup{bit}) \right\}
\end{align}
a more realistic description of an experiment is obtained.
For a well-tuned qubit the transversal lifetime is limited by the longitudinal since $T_2 \rightarrow 2T_1$ \cite{rasmussenSuperconductingCircuitCompanion2021,watsonAtomicallyEngineeredElectron2017}. By the discussion of \cref{sec:relax}, a factor on the order of unity in the relaxation rate does not matter much, hence we choose
\begin{align} \label{eq:phase_rate}
	\gamma_\textup{phase} &= \frac{\gamma_\textup{bit}}{2}.
\end{align}
From this point on, $\gamma_\textup{bit}$ will be referred to just as $\gamma$ since no confusion should arise.

Simulations analogous to \cref{fig:loschmidt_echo_noise} with the noise model from \cref{eq:real_noise} is seen in \cref{fig:real_noise}, which displays the noiseless and a realistic case along with one where noise drowns the target dynamics. Comparing \cref{fig:loschmidt_echo_noise,fig:real_noise} the realistic case shows a slight suppression of the Loschmidt echo from the bit-flip noise, but the noise rate is too small for phase decoherence to significantly disturb the structure and DQPTs are easily located through the minima of the Loschmidt echo. Remarkably, the $\gamma = \num{1e-1}$ case still displays a resemblance between the Loschmidt echo minima and the target zeros at early times, which then ultimately drowns in noise.
Additionally the panel shows both a clear suppression and distortion of the Loschmidt echo that can be traced to the two types of relaxation channels (\cref{sec:phase,sec:bit_flip}).

Using the qualitative pictures of the effect of phase decoherence (\cref{sec:phase}) and bit-flips (\cref{sec:bit_flip}) on the Loschmidt echo one can thus easily understand how and why an experimental result differs from the noiseless one. At a quantitative level structures are clearly preserved and we expect that DQPTs can reliably be studied in the realistic case.

\begin{figure*}
    \centering
    \includegraphics[width=\textwidth]{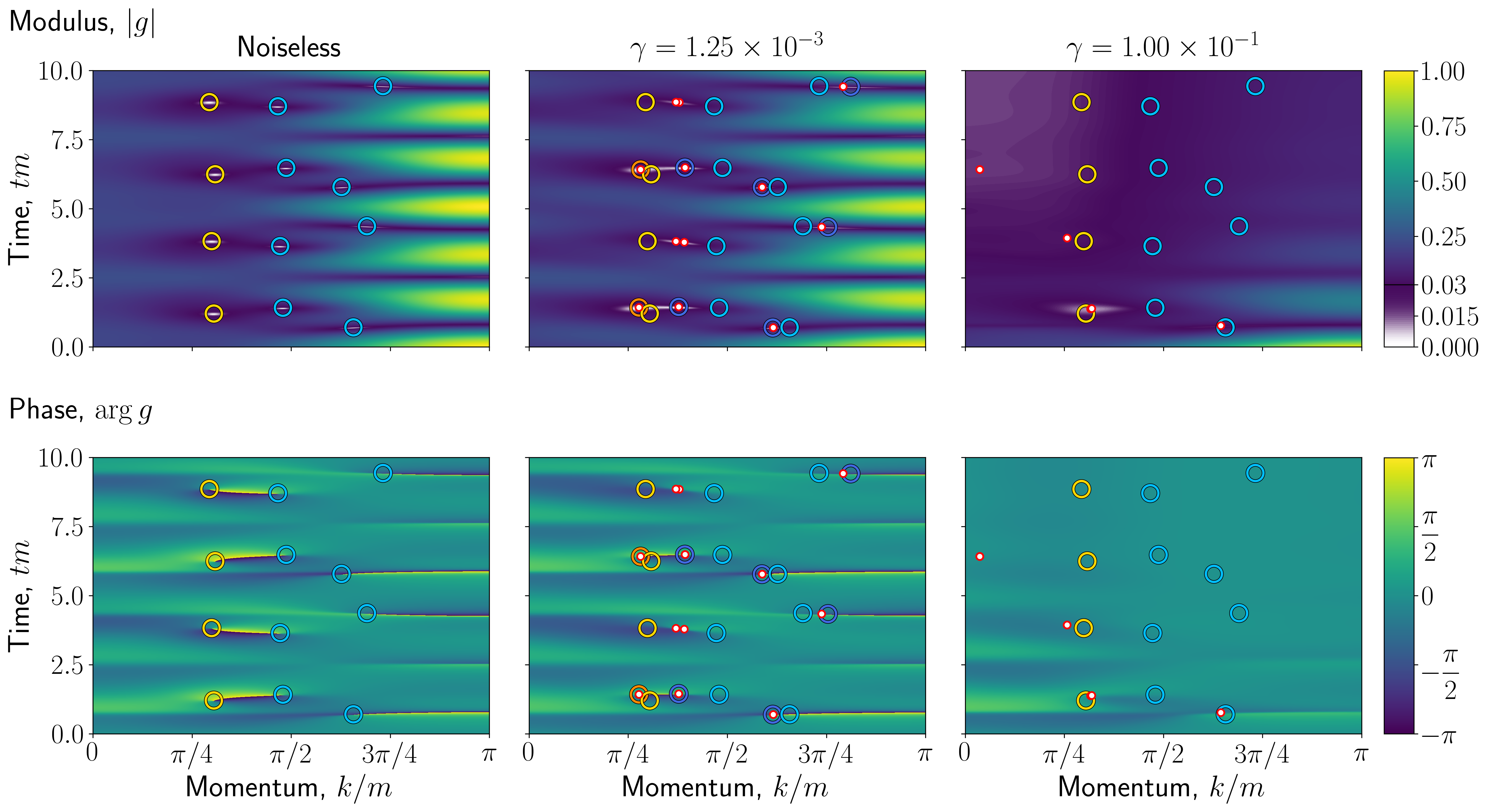}
	\caption{Simulations with the noise model in \cref{eq:real_noise} and the coupling constant fixed at $J/m = \num{1.95}$. The upper panels display the order parameter modulus and the lower panels the order parameter phase. Vortices in the phase of the order parameter in the noiseless case are marked with light blue and yellow circles for right and left wound vortices respectively. Vortices from the noisy simulations are marked with dark blue and orange circles for right and left wound vortices respectively. Additionally minima in the order parameter modulus are marked with red encircled white dots. The issue of the Loschmidt phase, as discussed in the context of \cref{eq:order_param_noise}, is clear, but the structure of tears and vortices is still discernible. Noticeably the time coordinate of the vortices is far more robust to the noise than the momentum coordinate.}
	\label{fig:order_param}
\end{figure*}

\subsection{Order parameter} \label{sec:order_param}
To study DQPTs, correlations between an order parameter and the Loschmidt echo is highly interesting. Specifically, a correlation between troughs in the Loschmidt echo and zeros of \cref{eq:order_param} was observed \cite{pedersenLatticeGaugeTheory2021}.

\subsubsection{Locating zeros}
To locate zeros of \cref{eq:order_param_noise} the following method is used. Consider a complex function of two real variables $f(x,y)$ expressed in polar form as $f = |f|\exp(i\phi)$.
At a zero of such a function the polar angle, $\phi$, is undefined, but apart from discontinuities of $2\pi$ it is smooth in the neighbourhood of that zero. There has to be a discontinuity since the phase has to traverse all possible angles as there would otherwise be a meaningful, smooth extension of the phase at the critical point.

Numerically, zeros on a possibly coarse grid of data points can thus be found by locating the ends of these lines of discontinuity. This method has the advantage of ensuring that, upon convergence, the zeros located are true zeros of the function and not just a point of small but non-zero modulus. This holds as long as discontinuities can reliably be discerned from rapid but continuous changes. For more details see \cite{pedersenLatticeGaugeTheory2021} and references therein.

\subsubsection{Simulation}
Choosing the noise model in \cref{eq:real_noise} and fixing the coupling constant at $J/m = \num{1.95}$, we obtain \cref{fig:order_param}.
Most of the vortices in the order parameter phase, and thereby zeros, from the ideal case can be located, but they are clearly more sensitive to the noise than minima in the Loschmidt echo. This is not surprising since the order parameter, \cref{eq:order_param_op}, couples all spins, whereas the Hamiltonian, \cref{eq:H}, only couple neighbouring spins. Noticeably, the momentum coordinate of the vortices is far more sensitive to the noise than the time coordinate. 
The robustness of the time coordinate indicates that the coherence between order parameter vortices and Loschmidt echo troughs from the noiseless case survives to the realistic, noisy case. For this objective it is, however, troublesome that some vortices have disappeared into the noise. Fortunately the missing vortices are still detectable as minima in the modulus which show the same robustness in the time coordinate as the vortices.

Due to the more complex nature of the order parameter tracing, the effect of the two noise types is less straightforward to analyze than was the case with Loschmidt echo. As the order parameter is gauge invariant, the main effect of the bit-flip channel should still be suppression of the modulus --- an effect seen in the middle panels of \cref{fig:order_param}. The precise impact of the phase channels is, however, difficult to discern.

\begin{figure*}
 	\centering
 	\includegraphics[width=.85\textwidth]{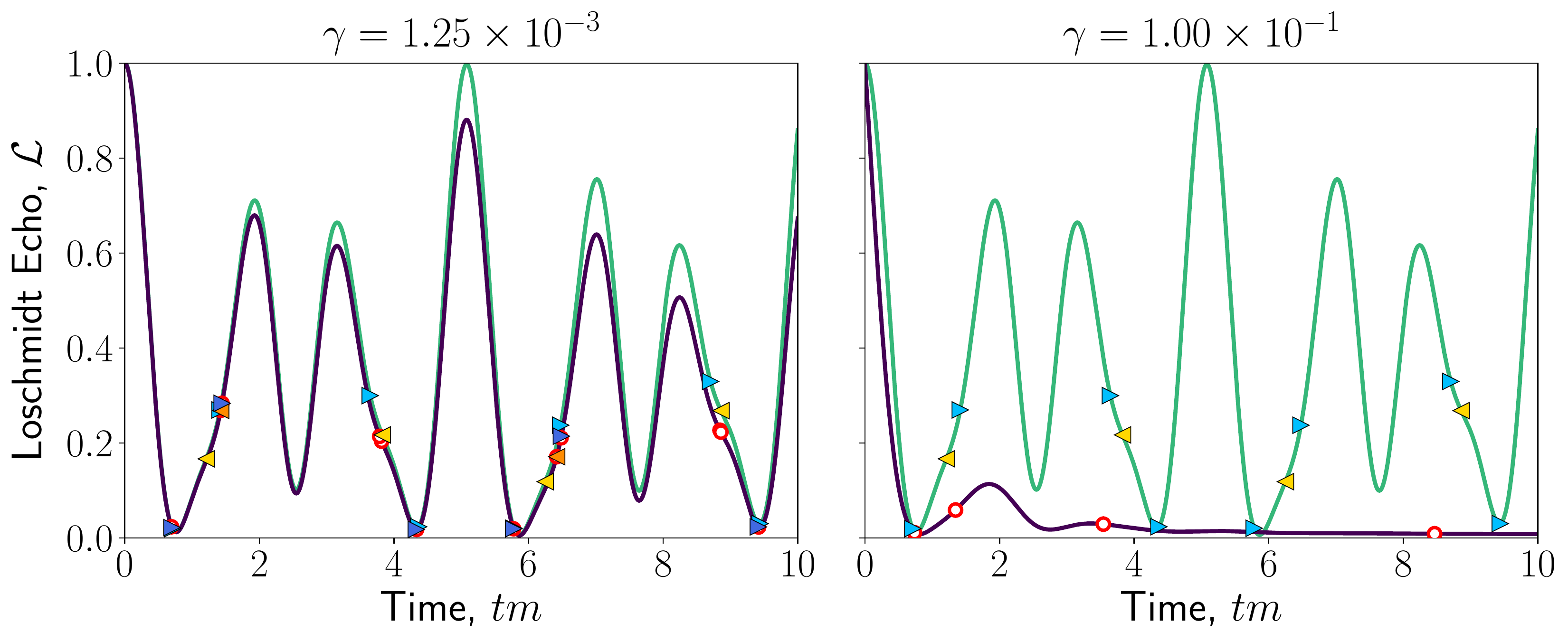}
	\caption{Loschmidt echo with the coupling constant fixed at $J/m = 1.95$, like \cref{fig:loschmidt_bit}, with the noise model of \cref{eq:real_noise} at the rates above each panel. The purple line in each panel shows the Loschmidt echo in the noisy case with a green line indicating the noiseless case. Zeroes of the order parameter are indicated by coloured triangles with the same colour code as in \cref{fig:order_param}. Minima of the order parameter are indicated by red encircled white dots. The previously established conclusions regarding the order parameter zeros, \cref{fig:order_param}, are seen here as well. As expected the robustness of the time coordinate allows good retention of the desired correlation. Additionally the missing zeros can be approximated by minima, which show a similar degree of correlation, but they are less reliable. This can be seen e.g. at $tm \sim 9$ where two minima have almost coalesced into the same point and thus looks like the same zero has been counted twice. There is no handedness to mark them as distinct.}
	\label{fig:loschmidt_order}
\end{figure*}
\subsection{Loschmidt echo order parameter correlation}
\begin{figure*}
    \centering
    \includegraphics[width=\textwidth]{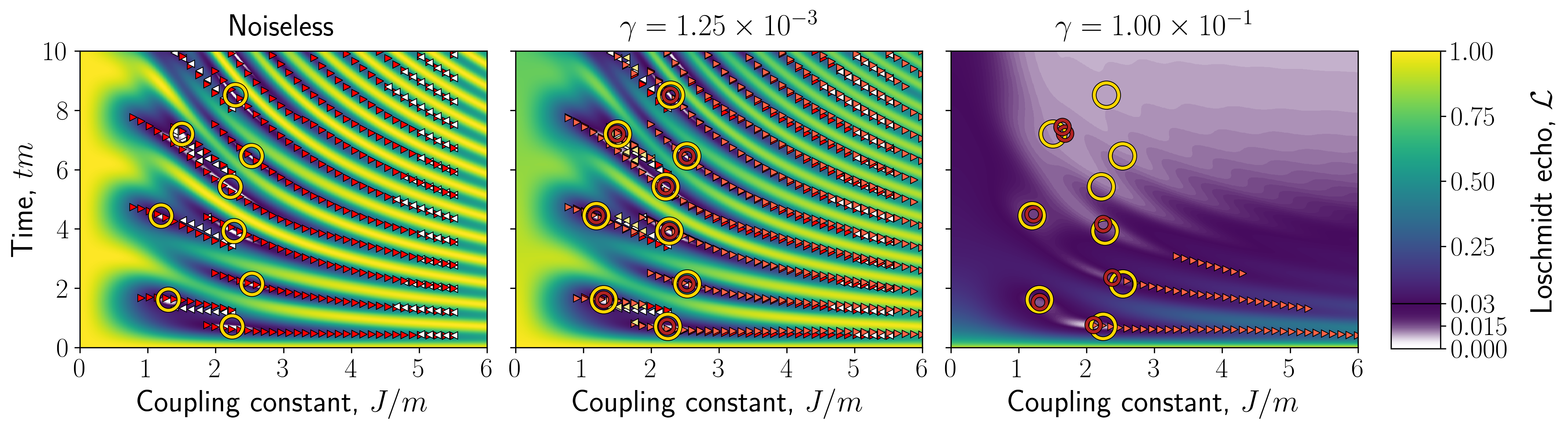}
	\caption{Loschmidt echo like in \cref{fig:real_noise} but this time with the addition of order parameter vortices. White arrowheads pointing to the left indicate left-wound vortices in the order parameter and red arrowheads pointing to the right right-wound vortices --- all from the noiseless simulation. In the noisy cases the colours are light yellow and a different shade of red respectively. The Loschmidt echo minima in the noisy cases are now marked by dark red circles and the zeros from the noiseless case larger yellow circles as the red encircled white dots are indistinguishable from the arrowhead. In the $\gamma = \num{1.25e-3}$ case the resemblance to the noiseless case is very good and the correlation between order parameter zeros Loschmidt echo valleys is retained. There are some subtle deviations and a few order parameter zeros are missing as seen from the red and white arrowheads popping up around $J = \num{2}$ in line with \cref{sec:order_param}. In the right-hand panel there is still some correlation and structure remaining but it is confined to early times. The Loschmidt echo is also seen to be slightly more noise resistant than the order parameter.}
	\label{fig:correlation}
\end{figure*}

\cref{fig:loschmidt_order} shows the Loschmidt echo from the simulations with the same parameters as \cref{fig:order_param}. In the noiseless case there is a correlation between order parameter zeros and valleys in the Loschmidt echo. The order parameter zeros do not seem to correlate with a Loschmidt echo minimum at $tm \sim 2$ and $8$ at this value of the coupling constant.
That is because order parameter zeros appear and move discontinuously as a function of $tm$ and $J/m$, whereas the Loschmidt echo is a continuous function of both. In \cref{fig:correlation} we will see that an order parameter zero will indeed show up in these minima at larger values of $J/m$. Likewise, the order parameter zeros at saddle points of the Loschmidt echo (e.g. $tm \sim 6.5$) are there because a valley in the Loschmidt echo exist close by in the $Jt$ plane.

As expected from \cref{sec:order_param}, the correlation is to a large extent retained in the noisy simulation at realistic relaxation rates. Noticeably, the pairs of order parameter zeros at the Loschmidt echo saddle points near $tm = 4$ and $tm=9$ have vanished. There are two order parameter minima there, which can still be detected, but they are very close. Without the vortex in the phase the handedness is lost, and lacking handedness and/or the noiseless simulation it would be difficult to distinguish such a pair of zeros from e.g over-counting of a single zero or experimental error, i.e. false positives and/or negatives might be an issue. Since the time coordinate is also significantly affected, minima in the order parameter modulus as an estimate of true zeros should be treated with caution as their reliability is somewhat questionable.

In the limit of large noise, both the Loschimdt echo and order parameter are so severally affected by the noise that any correlation between structures in the two is unlikely. No vortices in the phase of the order parameter remains and since the minima found responds more to the noise than any structure in the target function no correlation is seen.

Adding zeros of the order parameter as coloured arrowhead to \cref{fig:real_noise}, we obtain \cref{fig:correlation}. Red arrowheads poiting to the right indicate right handed vortices and vice versa for the white arrowheads in the noiseless case, and for the noisy case pale red and yellow triangles are used. The difference between the two cases is thus seen by the appearence of red and white triangles from beneath the pale colours. The minima in the Loschmidt echo are now marked by red circles, since dots would be difficult to distinguish from the arrowheads. Overall, the coherence from the noiseless case survives to the realistic case very well, but obviously not to the case with large noise. There are a few order parameter zeros missing and a slight deviation in the lines of zeros from the target, but these are minor details. Perhaps not surprisingly these deviations are largest around $J/m = 2$ where the most intricate dynamics occur. These dynamics are also the most interesting, but the differences are most pronounced at the end of lines and thus in some sense far from the Loschmidt echo zero in the corresponding valley and thereby the DQPTs of interest. 

\begin{figure*}
	\centering
	\includegraphics[width=\textwidth]{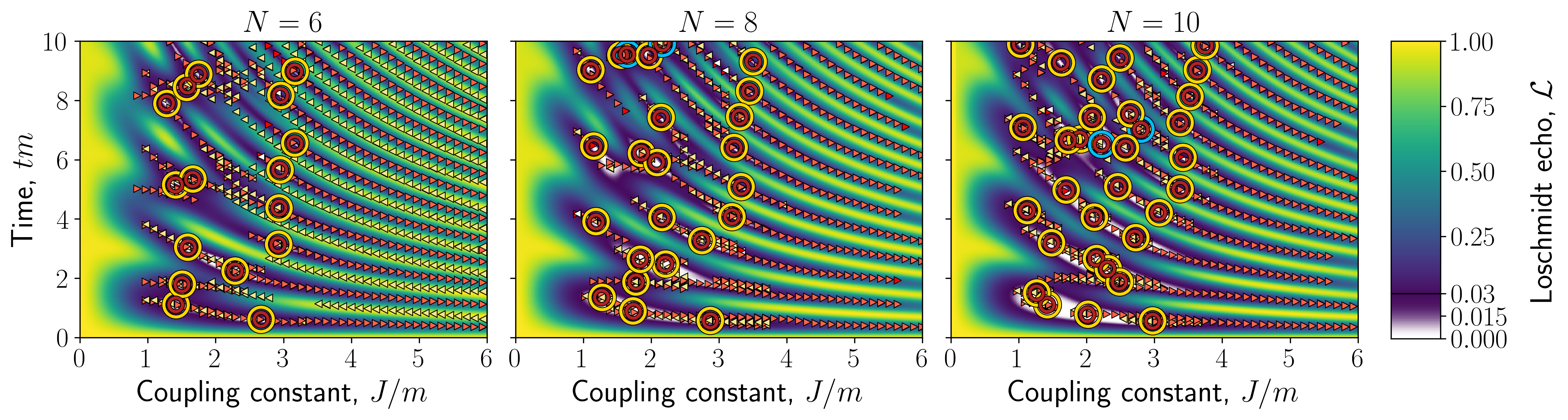}
	\caption{Loschmidt echo with minima and order parameter zeros analogous to the $\gamma = \num{1.25e-3}$ case from \cref{fig:correlation} for larger systems. The system size is displayed above each panel and the markers used are identical to \cref{fig:correlation}. At $N \geq 8$ one sees both left and right wound Loschmidt phase vortices in the noiseless case, hence these are marked with light blue and yellow for right and left respectively, like in \cref{fig:order_param}. For these simulations the noise model used was $\mathfrak{L}_\textup{phase}$, \cref{eq:phase_model}, with $\gamma = \num{8.33e-4}$ due to the size of the Hilbert space, $\dim\mathcal{H} = 2^{2N}$. It is however clear that the noisy simulations reproduce the unitary dynamics remarkably well at realistic noise rates with respect to the location of zeros of the Loschmidt echo and order parameter.}
	\label{fig:scale}
\end{figure*}
\subsection{Scaling} \label{sec:scaling}
Lastly it is important to consider how the system scales with $N$ and whether larger systems respond differently to noise. As the dimension of the full Hilbert space scales exponentially with the system size, simulations with non-gauge invariant relaxations channels are intractable with our code and these simulations thus use the noise model $\mathfrak{L}_\textup{phase}$ from \cref{eq:phase_model} with $\gamma = \num{8.33e-4}$. This relaxation rate corresponds to $\gamma_\textup{phase}$ from the noise model of \cref{eq:real_noise} at $\gamma_\textup{bit} = \num{1.25e-3}$ and $\gamma_\textup{phase} = \gamma_\textup{bit}/1.5$ --- a rate chosen to avoid unrealistically low noise impact due to the lack of bit flip channels. No major issues should, however, arise since the bit flip channels just suppress the Loschmidt echo at relaxation rates on this order of magnitude (\cref{sec:bit_flip}) largely without disturbing the structure. The results of such simulations for $N = 6, 8, 10$ are seen in \cref{fig:scale} where the same system of markers are used as in \cref{fig:correlation}. Even with the more intricate structures of the observables for the larger systems, no significant difference from the noiseless case that would affect the previous conclusions are seen. The high degree of correlation between order parameter zeros and Loschmidt echo valleys is retained and the Loschmidt echo minima approximate the true zeros with sufficient accuracy. The blue circles indicate that in the noiseless case at those zeros the vortex in the Loschmidt phase is now left-handed as opposed to the $N = 4$ case were all are right-handed. This information is lost in the noisy case, hence all minima are marked with the same colour. There is thus no reason to expect larger systems to introduce unexpected noise effects.

\section{Conclusion} \label{sec:conclusion}
The unanswered question of how the inevitable noise in an experimental realization might affect the study of DQPTs in a quantum link model version of the massive Schwinger model has here been addressed.
It has been shown that due to the gauge symmetry of the target dynamics bit-flip channels at realistic relaxation rates have the effect of suppressing the magnitude of the observables of interest largely without affecting their structures. In general, gauge-breaking relaxation channels do not seem troublesome for this kind of study. The phase channels are described by gauge invariant operators and thus more directly tamper with these structures. However, at realistic noise rates the effect is small enough such that the target dynamics remain clearly discernible.

Adapting the observables to the open quantum systems formalism introduced some challenges. Most noticeably, the lack of the Loschmidt phase forces one to rely on minimization to localize zeros and thereby DQPTs. This introduces the possibility of false positives and makes the otherwise interesting vortex dynamics impossible to study. Luckily, this does not seem to be an issue in the Loschmidt echo case and the order parameter has sufficient information to locate vortices in the phase. This does imply losing a few order parameter zeros but not anything critical to an experimental study.
All of the above also seem to scale nicely to larger systems although this is challenging to simulate.

In general this study supports the conclusion of Ref. \cite{pedersenLatticeGaugeTheory2021} that the framework of quantum link models allows the experimental realization of the $(1+1)$D U$(1)$ lattice gauge theory in NISQ-era devices. Specifically, the system is robust enough towards noise that the target dynamics are within reach of such devices. Due to the simplicity of the noise models, the results are relevant to multiple platforms. Furthermore, since the system is capable of handling noise at relaxation rates chosen to represent a NISQ-era device of somewhat poor lifetime, an experimental realization is likely to be within reach.

\section*{Acknowledgements}
The numerical results presented in this work were obtained at the Centre for Scientific Computing, Aarhus \url{http://phys.au.dk/forskning/cscaa/}.

\end{document}